\documentclass[twocolumn,showpacs,showkeys,prc,aps,superscriptaddress]{revtex4}
\usepackage{graphicx}

\newcommand{ \be }{\begin{equation}}
\newcommand{ \ee }{\end{equation}}
\newcommand{ \bea }{\begin{eqnarray}}
\newcommand{ \eea }{\end{eqnarray}}
\newcommand{ \la }{\langle}
\newcommand{ \ra }{\rangle}

\begin{document}

\title{Multiplicity fluctuations in Au+Au collisions at
$\sqrt{s_{NN}}=130$ GeV}

\affiliation{Argonne National Laboratory, Argonne, Illinois 60439}
\affiliation{Brookhaven National Laboratory, Upton, New York 11973}
\affiliation{University of Birmingham, Birmingham, United Kingdom}
\affiliation{University of California, Berkeley, California 94720}
\affiliation{University of California, Davis, California 95616}
\affiliation{University of California, Los Angeles, California 90095}
\affiliation{Carnegie Mellon University, Pittsburgh, Pennsylvania 15213}
\affiliation{Creighton University, Omaha, Nebraska 68178}
\affiliation{Nuclear Physics Institute AS CR, \v{R}e\v{z}/Prague, Czech Republic}
\affiliation{Laboratory for High Energy (JINR), Dubna, Russia}
\affiliation{Particle Physics Laboratory (JINR), Dubna, Russia}
\affiliation{University of Frankfurt, Frankfurt, Germany}
\affiliation{Indiana University, Bloomington, Indiana 47408}
\affiliation{Insitute  of Physics, Bhubaneswar 751005, India}
\affiliation{Institut de Recherches Subatomiques, Strasbourg, France}
\affiliation{University of Jammu, Jammu 180001, India}
\affiliation{Kent State University, Kent, Ohio 44242}
\affiliation{Lawrence Berkeley National Laboratory, Berkeley, California
94720}\affiliation{Max-Planck-Institut fuer Physik, Munich, Germany}
\affiliation{Michigan State University, East Lansing, Michigan 48824}
\affiliation{Moscow Engineering Physics Institute, Moscow Russia}
\affiliation{City College of New York, New York City, New York 10031}
\affiliation{NIKHEF, Amsterdam, The Netherlands}
\affiliation{Ohio State University, Columbus, Ohio 43210}
\affiliation{Panjab University, Chandigarh 160014, India}
\affiliation{Pennsylvania State University, University Park, Pennsylvania 16802}
\affiliation{Institute of High Energy Physics, Protvino, Russia}
\affiliation{Purdue University, West Lafayette, Indiana 47907}
\affiliation{University of Rajasthan, Jaipur 302004, India}
\affiliation{Rice University, Houston, Texas 77251}
\affiliation{Universidade de Sao Paulo, Sao Paulo, Brazil}
\affiliation{University of Science \& Technology of China, Anhui 230027, China}
\affiliation{Shanghai Institute of Nuclear Research, Shanghai 201800, P.R. China}
\affiliation{SUBATECH, Nantes, France}
\affiliation{Texas A \& M, College Station, Texas 77843}
\affiliation{University of Texas, Austin, Texas 78712}
\affiliation{Valparaiso University, Valparaiso, Indiana 46383}
\affiliation{Variable Energy Cyclotron Centre, Kolkata 700064, India}
\affiliation{Warsaw University of Technology, Warsaw, Poland}
\affiliation{University of Washington, Seattle, Washington 98195}
\affiliation{Wayne State University, Detroit, Michigan 48201}
\affiliation{Institute of Particle Physics, CCNU (HZNU), Wuhan, 430079 China}
\affiliation{Yale University, New Haven, Connecticut 06520}
\affiliation{University of Zagreb, Zagreb, HR-10002, Croatia}
\author{J.~Adams}\affiliation{University of Birmingham, Birmingham, United
Kingdom}
\author{C.~Adler}\affiliation{University of Frankfurt, Frankfurt, Germany}
\author{M.M.~Aggarwal}\affiliation{Panjab University, Chandigarh 160014, India}
\author{Z.~Ahammed}\affiliation{Purdue University, West Lafayette, Indiana 47907}
\author{J.~Amonett}\affiliation{Kent State University, Kent, Ohio 44242}
\author{B.D.~Anderson}\affiliation{Kent State University, Kent, Ohio 44242}
\author{M.~Anderson}\affiliation{University of California, Davis, California
95616}
\author{D.~Arkhipkin}\affiliation{Particle Physics Laboratory (JINR), Dubna,
Russia}
\author{G.S.~Averichev}\affiliation{Laboratory for High Energy (JINR), Dubna,
Russia}
\author{S.K.~Badyal}\affiliation{University of Jammu, Jammu 180001, India}
% ...

\collaboration{STAR Collaboration}\homepage{www.star.bnl.gov}\noaffiliation

\date{\today}
\begin{abstract}
We present the results of charged particle 
fluctuations measurements in Au + Au collisions at $\sqrt{s_{NN}}=130$ GeV
using the STAR detector. 
Dynamical fluctuations measurements are presented for inclusive 
charged particle multiplicities  as
well as for identified charged pions, kaons, and protons. 
The net charge dynamical fluctuations are found to be large and negative providing
clear evidence that positive and negative charged particle production is 
correlated within the pseudorapidity range investigated. Correlations are 
 smaller than expected based on model-dependent predictions for
a resonance gas or a quark gluon gas which undergoes fast hadronization and
freeze-out. Qualitative agreement is found with comparable scaled p+p measurements 
and a HIJING model calculation based on independent particle collisions, although a
small deviation from the $1/N$ scaling dependence expected from this model
is observed.
\end{abstract}

\pacs{PACS number: 25.75.Ld}
\keywords{Relativistic Heavy Ions, Event-by-event fluctuations.}
\maketitle

A key question of the heavy ion program at the Relativistic Heavy Ion
Collider (RHIC) is to understand whether the hot matter produced in the
midst of heavy ion collisions undergoes a transition to 
and from a quark gluon plasma (QGP) phase before it hadronizes.
One of the most striking signatures of such a QGP-HG (hadron gas) phase transition 
could be a strong modification
in the fluctuations of specific observables measured on a per collision basis,
i.e. event by event~\cite{Shuryak98,Rajagopal99,Stodolsky95,Asakawa00}. 
Most often discussed are mean transverse momentum fluctuations 
(temperature fluctuations)
and particle multiplicity fluctuations. 
For the latter, predictions range from enhanced multiplicity
fluctuations connected to the production of QGP droplets 
and nucleation processes in a first order QGP -- HG
phase transition, to a strong suppression of fluctuations as a consequence
of rapid freeze-out just after the phase transition ~\cite{Asakawa00,Jeon00}. 
In this case, final state values of
conserved quantities, such as net electric charge, baryon number, and
strangeness would not be strongly modified from their values in the
QGP stage. 
Due to the large difference in the degrees of freedom in the QGP
and HG phases, measured fluctuations, of the net electric charge in
particular, could be reduced by a factor ranging from 2 to 4 if a QGP is
 produced ~\cite{Asakawa00,Jeon00}. 
The frequency of production and size of QGP 
droplets may critically depend on the collision impact parameter. 
Central collisions are generally
expected to lead to larger and more frequent QGP droplet production. 
An increase in the size and production frequency of QGP droplets with
increasing collision centrality might then be signaled by a sudden 
change in the fluctuations of produced particles such 
as anti-protons and kaons~\cite{Gavin00}, as well as pions.

In this paper, we report on a  measurement of charged particle
multiplicity fluctuations as a function of collision centrality in Au + Au
collisions at an energy of $\sqrt{s_{NN}}=130$ GeV. 
We study event-by-event fluctuations of conserved quantities at
near-zero rapidity in the center-of-mass rest frame
(mid-rapidity). Specifically,  we discuss fluctuations in
the difference of the number of produced positively and negatively charged
particles (multiplicities) measured in a fixed rapidity
range, defined as ~\cite{Pruneau02} 
\be
\nu_{+-} = \la \left(\frac{N_+}{\la N_+\ra}
-\frac{N_-}{\la N_-\ra}\right)^2 \ra,
\ee
where $N_+$ and $N_-$ are multiplicities of positive and negative
particles calculated in a specific pseudorapidity, and transverse momentum range. 
The notation ``$\la O \ra$'' denotes an average of the quantity $O$ over an ensemble 
of events. The method used to calculate the averages $\la N_+\ra$ and $\la N_-\ra$,
which vary with collision centrality, is described in the following (see Eqs. \ref{nudynrecep1}-\ref{nudynrecep2}
).
We consider fluctuations in the production of all charged
particles, $N_+$ and $N_-$ (mostly pions) as well as specific cases
of proton and anti-proton, $N_p$ and $N_{\overline{p}}$, and positive 
and negative kaon, $N_{K+}$ and $N_{K-}$, fluctuations. 
The former amounts to a measurement of net electrical charge fluctuations, 
whereas the latter corresponds to measurements of net baryon number and net 
strangeness fluctuations. The method used to calculate this and other observables used
in this work is described in the following.

A difficulty inherent in the interpretation of measurements of multiplicity
fluctuations is the elimination of effects associated with
uncertainties in the collision centrality, often referred to as volume 
fluctuations. Event-by-event impact parameter variations, in particular, induce
positive correlations in particle production which do not depend on the
intrinsic dynamical properties of the colliding system, but rather simply 
reflect changes in the number of collision participants. 
Fluctuations in the difference of relative multiplicities, $\nu_{+-}$, 
defined in Eq.~1, are however free from this problem. 
This analysis is thus restricted to the study of such relative
multiplicities. As shown in ~\cite{Pruneau02}, $\nu_{+-}$ can be readily translated into 
observables $D$, and $\omega_Q$, discussed by other 
authors~\cite{Asakawa00,Jeon00,Gavin00}.
Its relation to the two-particle density is discussed below.
We will additionally study the behavior of relative multiplicities, $\nu_{+-}$,
 and other quantities of interest defined in this paper as a function of the
collision centrality estimated on the basis of the 
total charged particle  multiplicity measured in the pseudorapidity range $|\eta|<0.75$
in order to identify possible changes in the fluctuations with collision centrality.

The magnitude of the variance, $\nu_{+-}$, is determined by both
statistical and dynamical fluctuations.  Statistical fluctuations arise due 
to the finite number of particles measured, and can be readily calculated 
based on expectation values for Poisson distributions as follows:
\be
\nu_{+-,stat} = \frac{1}{\la N_+ \ra }+\frac{1}{\la N_- \ra}.
\ee
The statistical fluctuations depend on the 
experimental efficiency and analysis cuts used in the reconstruction 
of charged particle trajectories (tracks). 
The intrinsic or dynamical fluctuations are defined and evaluated as the
difference between the measured fluctuations and the statistical limit
\be
\nu_{+-,dyn} = \nu_{+-} - \nu_{+-,stat}. 
\ee
As shown in ~\cite{Pruneau02}, the dynamical fluctuations, $\nu_{+-,dyn}$, 
can be expressed as follows:
\be
\nu_{+-,dyn}=\bar{R}_{++}+\bar{R}_{--}-2\bar{R}_{+-},
\label{nudyn}
\ee
where  $\bar{R}_{ab}$ with $a,b=+,-$ are the averages of the 
correlation functions often used in multi-particle
production analysis~\cite{Foa75,Whitmore76,Boggild74}:
\be
\bar{R}_{ab}=\frac{\int_{\Delta\eta}R_{2,ab}(\eta_a,\eta_b) 
\rho_{1,a}(\eta_a)  \rho_{1,b}(\eta_b)d\eta_a d\eta_b}
   {\int_{\Delta\eta} \rho_{1,a}(\eta_a) d\eta_a
     \int_{\Delta\eta} \rho_{1,b}(\eta_b) d\eta_b  },
\label{Rab}
\ee
where  $R_{2,ab}=\rho_{2}(\eta_a,\eta_b)/(\rho_{1,a}(\eta_a)
 \rho_{1,b}(\eta_b)) -1$;
 $\rho_1(\eta)=dn/d\eta$, and  
$\rho_{2}(\eta_a,\eta_b)=d^2n/d\eta_a d\eta_b$
are single- and two-particle pseudorapidity densities respectively. 
The integrals could most generally be taken over the full particle phase space
($d^3p$) but are here restricted (without loss of generality) 
to pseudorapidity integrals to simplify the notation.
In cases where the produced particles are totally uncorrelated, 
two-particle densities
can be factorized as products of two single-particle densities. 
The correlators 
$\bar{R}_{ab}$ shall then vanish, and the measured dynamical fluctuations, 
$\nu_{+-,dyn}$, should be identically zero.
A deviation from zero thus should indicate correlations in particle
production. If correlations are due to production via many sub-collisions,
localized sources, 
or clusters, one should further expect the strength of 
the correlation to be finite but increasingly diluted with increased 
number of production clusters
or sub-collisions (hereafter called ``clusters'').
The correlators, $\bar{R}_{ab}$, will be inversely proportional to 
the multiplicity of clusters, and thus also inversely proportional to 
the total measured multiplicity of (charged) particles ~\cite{Pruneau02}.  
Measurements at the ISR and FNAL, have shown that charged particles 
have long range 
(differential) correlations dominated by a dependence on the 
relative rapidity of the correlated particles.  
One thus expects, as shown in ~\cite{Pruneau02}, 
that the functions, $\bar{R}_{ab}$, and $\nu_{+-,dyn}$ should vary 
slowly with the detector acceptance 
as long as the rapidity width of the acceptance
is smaller or of the order of the long range correlation width. 
This should however be experimentally verified by varying 
the acceptance used in the determination of $\nu_{+-,dyn}$.

Authors ~\cite{Mrowczynski02,Gazdzicki99} have suggested that if the reaction 
dynamics do not change with collision centrality, the measure 
$\Phi\approx \la N_{ch}\ra \nu_{dyn} /8$ 
(where $N_{ch}$ is the charged particle multiplicity in the rapidity range
considered) should be 
independent of the collision centrality. Conversely, a significant collision 
centrality dependence of $\Phi$ or related observables should hint at a change
 in the collision dynamics. We shall thus study 
the collision centrality dependence of 
both $\nu_{+-,dyn}$ and $\la N_{ch}\ra \nu_{+-,dyn}$.
The correlators, $\bar{R}_{ab}$, and $\nu_{+-,dyn}$, are robust 
variables: their measurements 
are independent of the average (global) detection efficiencies 
involved in the determination of 
multiplicities $N_+$ and $N_-$~\cite{Pruneau02}. 
The measurement of $\nu_{+-,dyn}$ thus 
does not require explicit efficiency corrections. Second order 
corrections are, in principle, needed to account for variations
of the detection efficiency through the fiducial acceptance. 
In the present study, we verified that the relative variation of the
detection efficiency (about 10\% in the transverse momentum region 
under study) results in a systematic uncertainty less than or
equal to the statistical error of the measured values.

The data presented are from minimum-bias and central trigger 
samples of Au + Au at $\sqrt{s_{NN}}=130$~GeV acquired by the 
STAR experiment during the first
operation of the Relativistic Heavy Ion Collider (summer 2000).
Detailed descriptions of the experiment
and the Time-Projection-Chamber (TPC) can be found elsewhere~\cite{Thomas99}.
In minimum bias mode, events were triggered by a coincidence between the
two Zero Degree Calorimeters (ZDCS) located +/- 18 m from the interaction
center and a minimum signal in the Central Trigger Barrel (CTB), which consists
of scintillator slats surrounding the TPC. The central trigger sample was acquired by 
requiring a higher multiplicity cut with the CTB corresponding to 15\% of the total hadronic
cross section.

In order to minimize the need for corrections to account for dependence of 
the detector acceptance and reconstruction efficiency on the vertex
position, the analysis reported here was restricted to events produced
within $\pm 0.70$ m of the center of the STAR TPC along the beam axis. 
In this range, the vertex finding efficiency is 100\% 
for collisions which result in charged particle multiplicities larger than 50 tracks 
in the TPC acceptance. It decreases to 60\% for events with fewer 
than 5 tracks from the primary vertex. We verified that the measurement of
$\nu_{+-,dyn}$ is insensitive to the vertex position by comparing 
values measured for different vertex cut ranges.
About 180000 minimum-bias, and 80000 central trigger 
events were used in this analysis after cuts.

The centrality of the collisions is estimated from the total charged
particle track multiplicity detected within the TPC in the pseudorapidity
range $|\eta|<0.75$. 
We use eight contiguous centrality bins based on the fraction
of triggered events: 6\%, 11\%, 18\%, 26\%, 34\%, 45\%, 58\%, and 84\%. 
The trigger efficiency is estimated to $94\pm 2$\%. 
The above fractions thus correspond to a constant increase in the fraction 
of the geometrical cross section which is sampled by each multiplicity bin.

Particle production is studied for both negative and positive hadrons 
over a transverse momentum range extending from  0.1 to 5 GeV/c,
and for pseudorapidity ranges from 
$|\eta|\leq 0.1$ to 1.0 in steps of 0.1 unit of pseudorapidity.
Good track quality is required by restricting the analysis 
to charge particle tracks producing more than 15 hits within the TPC. 
One additionally requires that more than  50\% 
of the hits be included in the final fit of the track. 

One uses the particle energy loss ($dE/dx$) measured with the TPC 
to identify the particles as pions, kaons, and protons (and their anti-particles). 
Particle identification proceeds on the basis of a parametrization 
of the mean ($\la E_{loss} \ra$) and width ($\sigma$)
of the average energy loss expected for electrons, 
pions, kaons, and protons as a function of their momentum. The analyses 
for pions, kaons, and protons are performed using momentum ranges 
$0.1 < p <0.6$ , $0.1 < p <0.6$, and $0.12 < p <0.7$ GeV/c
respectively.  Lower bounds are set near or
below detection threshold to maximize particle yields. Upper bounds are used
to minimize cross species contamination. The inclusive analysis of all charged species 
is performed within the range $0.1 < p <5.0$ GeV/c. 
Limiting the particle momenta for this analysis to less than 
5 GeV/c insured that particle charge was not mis-assigned 
while allowing for a fully inclusive measurement
of the soft particle spectra.
Given that the bulk of the particle production
is below 2 GeV/c, the inclusive analysis is rather insensitive to the exact value of the upper bound which is used.
The detection efficiency rises from
zero to roughly 85\% within an interval of 0.1 GeV/c above detection thresholds, remaining 
constant for larger momenta.
Measured particles are tagged as pions if their measured energy loss
deviates by less than two standard deviations ($2\sigma$) from the expected mean for
pions of the same momentum, 
while deviating by more than $2\sigma$ for kaons of that same momentum.
Similarly particles are identified as kaons (protons) if the deviation
from the kaon (proton) mean energy is less than  $2\sigma$ while being 
larger than  $2\sigma$ from the pion and proton (kaon) mean energy loss. 
Contamination of the kaons and protons by pions
is negligible at low momentum, and  
estimated to be less than 5\% at the highest momenta accepted for
those particles. For cross-species contamination at this level, it 
was verified that the measurement is insensitive to the actual
value of the momentum cuts.

To reduce contamination from secondary electron tracks, 
and focus this analysis on primary tracks, i.e. particles produced 
at the Au+Au collision vertex, only  tracks which passed within 3~cm 
of the collision vertex were accepted. We verified electron (positron) contamination
has a negligible impact on our measurements of $\nu_{+-,dyn}$ by repeating the
analysis with and without an electron/positron exclusion cut based on the track
energy loss measured in the TPC, i.e. accepting tracks with a
 dE/dx more than 2 standard deviations away from the expected value for an
electron of the measured momentum.  

As already mentioned, the measurement of  $\nu_{+-,dyn}$ is
independent of the average detector efficiency. 
It is therefore also insensitive to particle losses, 
e.g. anti-protons, due to scattering through the detector. 
It is however sensitive, in principle, to the generation of 
background particles within the detector. 
The effect of such background particles (e.g. protons scattered off 
the beam pipe) is minimized by using the 3 cm distance of closest 
approach cut mentioned above.
Also, it was considered whether finite track splitting, 
possibly encountered in the reconstruction of charged particle 
tracks in the TPC,  may produce measurable effects on  $\nu_{+-,dyn}$. 
We verified that, within statistical errors,  the same value 
is obtained when the pseudorapidity regions used to count positive and
negative tracks were separated by a $\Delta\eta=0.25$ gap.

Since finite width multiplicity bins were used for this analysis,
values of $\nu_{+-,dyn}$ are multiplicity-bin averaged according to the
following expression:

\be
 \nu_{+-,dyn}(M_{low}\leq M<M_{high})=\frac{\sum \nu_{+-,dyn}(M)P(M)}{\sum P(M)}
 \label{nudynrecep1}
\ee
where $P(M)$ is the probability of having a total charge multiplicity, $M$,
and $\nu_{+-,dyn}(M)$ is given by
\bea
 \nu_{+-,dyn}(M)&=&\frac{\la N_+(N_+-1)\ra_M}{\la N_+\ra_M^2}\\ \nonumber
                &+&\frac{\la N_-(N_--1)\ra_M}{\la N_-\ra_M^2}\\ \nonumber
                &-&2\frac{\la N_+N_-\ra_M}{\la N_+\ra_M\la N_-\ra_M}
\eea
The notation $<O>_M$ is used to indicate the average of the quantity $O$ 
for all events with a charged particle multiplicity, M, in the pseudorapidity range $|\eta|<0.75$. 
Our analysis proceeds in two passes. The first pass involves the determination of 
the averages $\la N_{\pm}\ra_M$ as a function of the multiplicity, $M$, using unity bin-width in $M$
while the second pass uses these averages as coefficients in the above expression of $\nu_{+-,dyn}(M)$.
The averages $\la N_{\pm} \ra_M$ are determined from the events with multiplicity $M$:
\be
\la N_{\pm} \ra_M = \frac{1}{N_{ev}(M)} \sum N_{\pm}
\ee
The sum is taken over the $N_{ev}(M)$ events of multiplicity, $M$, present in our sample. 
The averages $\la N_{\pm}\ra_M$ thus obtained display a scatter determined by the finite statistics about a monotonically 
increasing trend (with M).  If uncorrected, this scatter, may induce an artificial change of the
value of $\nu_{+-,dyn}{\scriptstyle (M)}$ in each bin. To minimize this effect, we model (fit) the average
 $\la N_{\pm}\ra_M$ dependence on the multiplicity M with a polynomial optimized to reproduce 
 the shape of the dependence. We then determine
  $\nu_{+-,dyn}(M)$ using the averages 
$\la N_{\pm}\ra_{fit,M} \equiv {\overline N}_{\pm,M}$
 predicted by the fit rather than the actual
averages. The calculation of $\nu_{+-,dyn}$ in a finite width multiplicity bin then proceeds with the 
following expression:
\bea
 &\nu_{+-,dyn}&{\scriptstyle (M_{low}\leq M<M_{high})}=\\ \nonumber
& &	\frac{1}{N_{ev}}
	\sum_{events}  \left[\frac{N_+(N_+-1)} {{\overline N}_{+,M}^2}  \right. \\ \nonumber
                & &+\frac{N_-(N_--1)}{{\overline N}_{-,M}^2}\\ \nonumber
                & &\left.-2\frac{N_+N_-}{{\overline N}_{+,M} {\overline N}_{-,M}}\right] 
\eea
where the sum is taken over the $N_{ev}$ events in the multiplicity bin $M_{low}\leq M<M_{high}$.

The quantity $\la N\ra\nu_{+-,dyn}$ is determined 
in a similar fashion using the following expression:
\bea
 \label{nudynrecep2}
 &\la N\ra\nu_{+-,dyn}&{\scriptstyle (M_{low}\leq M<M_{high})}=\\ \nonumber
& &	\frac{1}{N_{ev}}
	\sum_{events} ({\overline N}_{+,M} +{\overline N}_{-,M} ) \\ \nonumber
& &	\left[\frac{N_+(N_+-1)} {{\overline N}_{+,M}^2}  \right. \\ \nonumber
                & &+\frac{N_-(N_--1)}{{\overline N}_{-,M}^2}\\ \nonumber
                & &\left.-2\frac{N_+N_-}{{\overline N}_{+,M} {\overline N}_{-,M}}\right] 
\eea
To study the effect of this method of bin-averaging, a simulation
was performed using HIJING events, comparing the results
of equation 10, and equation 3 in the limit of large statistics.
The HIJING model does not incorporate re-scattering and
should not therefore exhibit a significant centrality dependence.
The results showed that for all bins except the lowest multiplicity  bin
used for this analysis, the two equations gave the same result
within the quoted systematics. In the first multiplicity bin,
equation 10 yielded a result approximately 15\% larger than
equation 3.

\begin{figure}
\centerline{\includegraphics[width=3.9in]{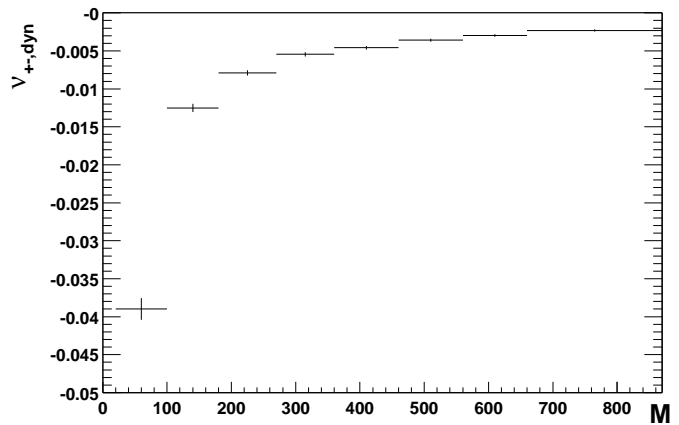}}
\centerline{(a)}
\centerline{\includegraphics[width=3.8in]{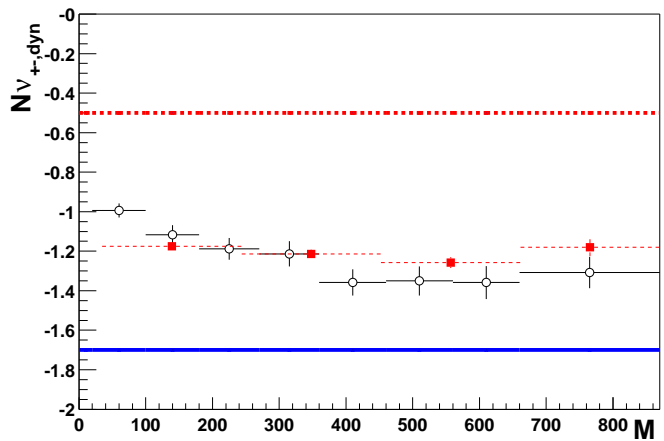}}
\centerline{(b)}	
\caption[]{(a) Dynamical fluctuations, $\nu_{+-,dyn}$, 
measured in $|\eta| \leq 0.5$
as a function of the collision centrality estimated 
with the total (uncorrected) multiplicity, $M$, in $|\eta|<0.75$. 
(b) $\la N\ra \nu_{+-,dyn}$ measured in $|\eta| \leq 0.5$ vs $M$ (opened circles) compared to the 
charge conservation limit (dotted line), resonance gas expectation based on
ref. ~\cite{Jeon00}(solid line); and HIJING calculation (solid squares)}
\label{nu2}
\end{figure}

Fig. \ref{nu2}a shows the dynamical fluctuations, $\nu_{+-,dyn}$, 
of the net charge measured in the pseudorapidity range $|\eta| \leq 0.5$, 
as a function of the total multiplicity, $M$,  
measured in the pseudorapidity range $|\eta| \leq 0.75$. 
The horizontal bars on the data points reflect the width of the multiplicity 
bins used in this analysis while the vertical bars reflect statistical
errors. 
We estimate the systematic errors based on data taken and analyzed with 
different trigger and analysis cuts, to be of the order of 2\%. 
An additional systematic uncertainty of the order of 3\% is derived
by a separate analysis of different data subsets. 
The dynamical fluctuations of the 5\% most central collisions
then amount to 
$\nu_{+-,dyn}=-0.00236 \pm 0.00006$(stat) $\pm 0.00012$(syst).
The dynamical fluctuations are finite and negative: a clear 
indication that positive and negative particle production
are correlated within the pseudorapidity range  
considered (see Eq. \ref{nudyn}). 
One observes the strength of the dynamical fluctuations 
decreases monotonically with increasing collision centrality. This can
be understood from the fact that more central Au+Au collisions involve
an increasing number of ``sub-collisions'' (e.g. nucleon-nucleon
collisions): the two-particle correlations are thus increasingly 
diluted and the magnitude of $\nu_{+-,dyn}$ is effectively reduced.

We compare our results, for the most central collisions,
 to those recently reported by the PHENIX collaboration~\cite{Phenix02}
which measured  net charge fluctuations in terms of the relative
variance $\omega_Q = \la \Delta Q^2\ra/N_{ch}$ in the rapidity range
 $|\eta|<0.35$, and the angular range $\Delta \Phi=\pi/2$, 
 for $p_{\perp}>200$MeV/c. 
They reported a value $\omega_Q=0.965\pm 0.007$(stat) $-0.019$(syst)
for the 10\% most central collisions.
The large (unidirectional) systematic error is reported 
to result from uncertainty in the correction applied for effects of finite detector efficiencies. 
In order to compare the PHENIX result with the present study, we use the expression in reference  ~\cite{Pruneau02}
\be
\nu_{+-,dyn}=\frac{4}{N_++N_-}\left( \omega_Q -1 \right),
\ee
The charged particle multiplicity in the PHENIX
detector acceptance is $79\pm 5$ for the 10\% most central
collisions.  This comparison gives $\nu_{+-,dyn}=-0.0018 \pm 0.0004\;(stat) - 0.009\;(syst)$
in agreement with the
value of  $\nu_{+-,dyn}=-0.00263 \pm 0.00009$(stat) $\pm
0.00012$(syst) we measure for 11\% central collisions.  
The agreement is best if one considers
the low bound of the PHENIX measurement  which is maximally
corrected for finite efficiency (which is reflected in the systematic
error). The difference between the two results might 
be due, in part, to dependence of the multiplicity fluctuations on 
rapidity and azimuthal angle as well as acceptance effects.

It is important to consider the effects of charge conservation on 
the net charge fluctuations since they are expected to be non-negligible 
 even for small finite rapidity coverage ~\cite{Pruneau02}. 
The contribution is estimated to  be $-4/\la N\ra_{4\pi}$ 
where $\la N\ra_{4\pi}$ is 
the total number of charged particles {\em produced} by the
collisions.  
The PHOBOS collaboration has reported ~\cite{Phobos01} that the total charged 
particle multiplicity amounts to $4200 \pm 470$ in the
6\% most central Au+Au collisions at $\sqrt{s_{NN}} = 130$ GeV.
The charge conservation contribution to the measured dynamical 
fluctuations is thus of the order of $-0.00095\pm 0.0001$, i.e. 40\%
of the observed dynamical fluctuations. 

We next discuss the centrality dependence of the fluctuations. 
In central collisions, the measured dynamical fluctuations,
$\nu_{+-,dyn}$ are expected to be reduced due to
 dilution of the two-particle correlations. 
One expects the magnitude of $\nu_{+-,dyn}$ 
should scale inversely to the number of sub-collisions producing
particles. 
Assuming the average number of particles produced by such
sub-collisions is independent of the collision centrality, 
one then expects the fluctuations to scale inversely
as the charged particle multiplicity. 
The quantity  $\la N\ra\nu_{+-,dyn}$ should therefore be 
independent of collision centrality if no significant variation 
in the mechanism of the particle production 
arises with collision centrality. 
This notion was suggested by Gazdzicki ~\cite{Gazdzicki99} 
and Mrowczynski ~\cite{Mrowczynski02} in terms 
of the fluctuation measure $\Phi$ which, as shown in ~\cite{Pruneau02},
is equal to $\la N\ra\nu_{+-,dyn}/8$ for $\la N_+\ra \approx \la N_-\ra$. 
Fig. \ref{nu2}b shows the measured centrality dependence 
of  $\la N\ra\nu_{+-,dyn}$, calculated with Eq.~\ref{nudynrecep2}, for all charged particles produced in the pseudorapidity range $|\eta|\leq 0.5$. 
In this Figure, the charged particle multiplicity, $N$, is corrected 
for finite detection efficiencies using correction factors 
which depend linearly on the charged particle multiplicity (TPC detector occupancy)
with values ranging from 85\% to 70\% for peripheral and central collisions 
respectively 
~\cite{Star01}.
The measured values range from $-1$ to $-1.4$ and are approximately a factor of two larger
 than the charge conservation limit, shown as
a dotted line, in Fig. \ref{nu2}b). This indicates dynamical fluctuations are not only finite 
but in fact rather large.  As discussed in detail below, the values 
measured for $\la N\ra\nu_{+-,dyn}$ however fall
short of predictions for a resonance gas in equilibrium (approximately $-1.7$; solid line)
and for a scenario involving a quark-gluon gas undergoing fast hadronization (approximately -3.5; not shown in Fig. \ref{nu2}b) ~\cite{Jeon00}.
The measured values are in qualitative agreement with a calculation based on 
HIJING (solid squares) ~\cite{Wang01}.
Indeed, the values predicted by HIJING are within 20\% of the measured values at all centralities. 
While the HIJING calculation is independent of collision centrality,
the experimental data exhibit a small but finite centrality dependence which is significant above 
the first bin in Fig. 1b. The HIJING calculation does not feature
rescattering, and is therefore not expected to exhibit a significant centrality dependence.
The observed centrality dependence may then suggest there are rescattering effects, or other 
dynamical effects with centrality, and its interpretation requires further investigation.

\begin{figure}
\centerline{\includegraphics[width=3.3in]{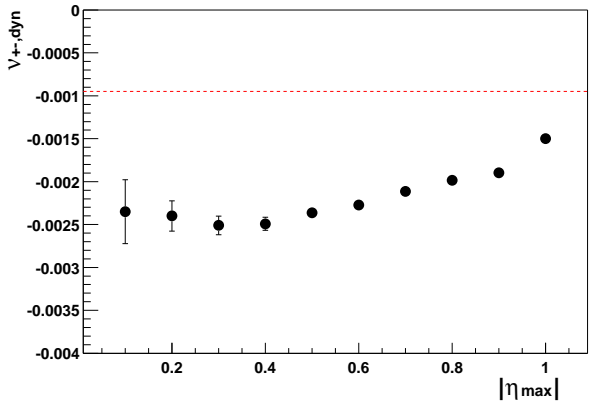}}
\caption[]{ Fluctuations $\nu_{+-,dyn}$ for the 6\% most central 
collisions as a function of the range of integrated pseudorapidities. 
The expected limit due to charge conservation is shown as a dotted line. }
\label{mNuVsEta}
\end{figure}

The magnitude of the net charge dynamical fluctuations is determined 
by the strength of the two-particle correlations in the integrated 
rapidity range. 
Measurements from p+p collisions at the ISR and $p+\overline{p}$ collisions at FNAL indicate  that the relevant
rapidity interval for two-particle correlations is approximately one unit. 
One thus expects the 
dynamical fluctuations to exhibit a mild dependence on the rapidity 
range used for the measurement ~\cite{Pruneau02}. 
Fig. \ref{mNuVsEta} shows the measured dynamical fluctuations (filled circles) 
as a function of the pseudorapidity range.
The pseudorapidity integration range is varied from $-0.1<\eta <0.1$ 
to  $-1.0<\eta <1.0$ in discrete steps of 0.1 units of
pseudorapidity. 
Error bars shown are statistical only. 
Focusing on the region in Fig. 2 where systematic effects
due to finite  are expected to be small, we examine the
data for $\eta > 0.4$. One observes the absolute value of the the
dynamical fluctuations is largest in this range for
$|\eta| \approx 0.4$, and that it decreases monotonically
for larger acceptance without, however, reaching the charge
conservation limit. One finds $|\nu_{dyn}|$ decreases by
35-40\% while the integrated pseudorapidity range is
increased by a factor of 5 from 0.4 to 2 pseudorapidity
units. The dependence of dynamical fluctuations on
the experimental acceptance is rather modest. 
In contrast, the $\Phi$ measure increases approximately by a factor of 10
from $-0.1<\eta <0.1$ to  $-1.0<\eta <1.0$ due to its explicit 
dependence on the pseudorapidity bin size.

We next consider the above  results in the light of correlation functions 
measured in $p+p$ and $p+\bar{p}$ collisions at CERN and FNAL ~\cite{UA5-88,Amendolia76,Foa75}
with the use of Eq. \ref{nudyn}. 
To account for the unavailability of p+p comparison data at the same energy as RHIC, an interpolation was
made using results obtained at lower and higher collision 
energies (parameterization from~\cite{UA5-86}). 
Based on results published in Refs~\cite{UA5-88,Amendolia76,Foa75}, we also note that the correlation function for oppositely charged 
particles, $R_{+-}(y_+ \approx y_-)$, 
is found to be approximately twice as strong 
as the same sign particles correlations, 
$R_{++}\approx R_{--}$~\cite{Foa75,Whitmore76}, 
and that it is independent of the collision energy.
The CERN and FNAL measurements~\cite{UA5-88,Amendolia76,Foa75} find 
the single charged particle and two-particle (charged-charged) pseudorapidity 
densities  to be respectively $\rho_1(\eta=0)\approx 2.06$  and 
$C_2(0,0)=\rho_2(\eta_1=0,\eta_2=0)-\rho_1(\eta_1=0) \rho_1(\eta_2=0)
\approx 2.8$. The charged-charged correlation integral $R_{cc} = (R_{++}+R_{--}+2R_{+-})/4$ 
 is thus $R_{cc}\approx 0.66$ (see ref. ~\cite{Pruneau02}). 
Furthermore, assuming equal multiplicities of positively and 
negatively charged particles, one finds for the charged-charged correlation
$R_{cc}\approx  1.5 R_{++}$, 
which we use to estimate the correlation measured in this work as 
$\bar{R}_{++}+\bar{R}_{--}-2\bar{R}_{+-}\approx -2 \bar{R}_{++} 
\approx 4\bar{R}_{cc}/3 \approx 0.88$. 
The pseudorapidity densities are very different in p+p and A+A
collisions. 
Under assumption that the correlations are due to production in a 
finite number of sources (clusters), they should be inversely 
proportional to the particle density.
In the 5\% most central Au+Au collisions, the pseudorapidity charged particle density ($dN/d\eta$) is 
about $526\pm 2(stat)\pm 36 (syst)$
~\cite{Star01} compared to approximately 2.06 in $p\bar{p}$ collisions. 
Such a dilution would give for the correlation function 
a value of $0.88 \cdot 2.06/526 \approx 0.0034$, in qualitative agreement
with the measured values for Au+Au collisions presented in this paper. 
We stress that valuable insight can be gained by comparing the current
130 GeV data and upcoming 200 GeV Au+Au analysis with explicit measurements made
in p+p collisions rather than using the above first order approximation.

We next compare our measurement of the dynamical fluctuations
to predictions of net charge fluctuations based on thermal models 
~\cite{Asakawa00,  Jeon00,Jeon99, Koch02, Bleicher00}.
To this end, we express our measurement of $\nu_{+-,dyn}$ 
in the range $|\eta|\leq 0.5$ in terms of the $D$ variable 
introduced in ~\cite{Jeon00}, using
\be
D=4+\la N\ra\nu_{+-,dyn}
\ee
 valid for $N_+\approx N_-$~\cite{Pruneau02}. 
We find using data shown in Fig. \ref{nu2}b
 that $D$ decreases from $3.1\pm 0.05$ (statistical error only) for the most peripheral 
collisions measured to $2.8\pm 0.05$ in central collisions. 
However, a comparison to thermal model predictions 
requires the data to be corrected for charge conservation effects. 
One must subtract the charge conservation contribution which 
amounts to $\Delta D = -0.00095\times 526=-0.50 \pm 0.06$. 
The corrected values of $D$ thus range from $3.6\pm 0.1$ to $3.2\pm 0.1$. 
According to the discussion of Refs.~\cite{Asakawa00,  Jeon00,Jeon99, Koch02,Bleicher00}, 
these values approach that ($D \approx 2.8$) expected for a resonance gas. They 
are significantly larger than expected in the above referenced work ~\cite{Jeon99,Jeon00,Bleicher00,Koch02}
for a quark-gluon gas undergoing fast hadronization and freeze-out ($D \approx 1$). It is not
possible to draw a firm conclusion concerning the existence or non-existence of a 
deconfined phase during the collisions from these results since,
as the above authors have pointed out, incomplete thermalization could lead
to larger fluctuations than expected for a QGP. Other work  ~\cite{Longacre02} has
also suggested that the prediction of $D \approx 1$ for a quark-gluon gas is model
dependent, and that other effects such as gluon fragmentation prior to
to hadronization could increase the fluctuations expected even if a quark-gluon 
plasma were produced.

We extend the study of net charge fluctuations 
to identified particles and consider measurements of the net charge fluctuations of pions, 
kaons, and protons/anti-protons. 
Measurement of the $K^+$,$K^-$ and $p$,$\overline{p}$ net charge are of 
particular interest as they
address respectively fluctuations of net strangeness and baryon number which 
might be more sensitive to the details of the collision process. 
The results are compiled in Table 1 for all charged species, pions, 
kaons, and $p$,$\overline{p}$.
The results indicate that the dynamical fluctuations for pions are approximately 
the same magnitude as for inclusive non-identified charged particles.
One however discerns a small but finite difference, especially
for integrated pseudorapidity ranges $|\eta| \leq 0.7$ and larger.
The measurement of $K^+$,$K^-$ and $p$,$\overline{p}$ fluctuations 
is hampered by the smaller multiplicities and finite 
detection efficiencies for kaons and protons and their anti-particles. 
Our measurement, which is presented in Table~1 for acceptances 
from $|\eta| \leq 0.5$ to $|\eta| \leq 1.0$ is thus limited to a 
central collision trigger sample. The effect of the variation of efficiency 
near detection threshold was studied by changing the transverse momentum
threshold used in the determination of $\nu_{+-,dyn}$. It was found that for inclusive non-identified
particles, 
$\nu_{+-,dyn}$ changed by less than 3\% while varying the transverse momentum cutoff for particle
detection from 0.1 to 0.2 GeV/c. The same study using HIJING events led to a 10\%
change in $\nu_{+-,dyn}$. 

The systematic error for protons (anti-protons) is difficult to assess, since GEANT studies
indicate a considerable fraction of the proton yield below 0.4 GeV/c is associated with pion-induced 
proton knockout reactions in the beam pipe. Background protons 
bear little correlation with anti-protons. The terms $R_{++}$ and $R_{+-}$ involved in the
calculation of  $\nu_{+-,dyn}$ should have a Poissonian behavior, and therefore the 
contribution of uncorrelated background to these terms should partly cancel.
We find the value of $\nu_{+-,dyn}$ exhibit changes smaller than
the statistical uncertainties when raising the threshold from 0.2 to 0.3 GeV/c, and
hence ascribe a systematic error of the order of 20\% for the $p$,${\overline p}$ measurement.

The dynamical fluctuations of the charged kaons 
and $p$,$\overline{p}$ are also finite. 
Their size (absolute value) are in fact larger than the dynamical 
fluctuations measured for pions and for inclusive non-identified charged particles.
The proton dynamical fluctuations are somewhat larger than the kaon fluctuations.  
Strangeness conservation, and baryon number conservation
should influence the size of the dynamical fluctuations for the net charge of 
kaons, and $p$,$\overline{p}$ respectively. 
The charge conservation limit derived for inclusive non-identified charged particles
can be readily reinterpreted to estimate the
expected magnitude of dynamical fluctuations for $K^+$,$K^-$ and $p$,$\overline{p}$. 
One finds that the kaon and $p$,$\overline{p}$ dynamical fluctuations are similar 
or slightly larger than their respective charge conservation limits.

\begin{table}[ht]
\caption{$1000\times\nu_{+-,dyn}$, 
for charge pions, kaons, and protons, as a function
of the integrated pseudorapidity range}
\begin{tabular}{|c|c|c|c|c|}
\hline
 $|\eta|$ & All +-& $\pi^{\pm}$ & $K^{\pm}$ & $p$,$\overline{p}$\\
\hline 
\hline
0.5 &$-2.36\pm 0.06$&$-2.4\pm 0.1$& $-5\pm 3$   & $-3\pm 7$  \\   
0.6 &$-2.27\pm 0.04$&$-2.4\pm 0.1$& $-5\pm 2$   & $-5\pm 3$  \\   
0.7 &$-2.11\pm 0.04$&$-2.18\pm 0.08$& $-4\pm 2$   & $-7\pm 5$  \\
0.8 &$-1.98\pm 0.03$&$-2.12\pm 0.07$& $-6\pm 2$   & $-8\pm 3$  \\
0.9 &$-1.90\pm 0.03$&$-2.02\pm 0.06$& $-6\pm 2$   & $-9\pm 2$  \\
1.0 &$-1.75\pm 0.02$&$-1.92\pm 0.06$& $-7\pm 1$   & $-8\pm 2$  \\
\hline
\end{tabular}
\normalsize
\label{tab:nudyn}
\end{table}

{\bf Summary and Conclusions}

We have measured event-by-event net charge dynamical fluctuations 
for inclusive non-identified charged particles, as well as for 
identified pions, kaons, and protons and their anti-particles in 
Au+Au  collisions at  $\sqrt{s_{NN}}=130$ GeV. Dynamical fluctuations measured for
 inclusive non-identified charged particles
are finite and exceed by nearly a factor of two expectations based on charge conservation. 
 We find the magnitude of the net charge dynamical fluctuations 
to be in qualitative agreement with expectations based on measurements of
charged particle correlation functions in p+p collisions measured at
the ISR. We however find that although 
the fluctuations roughly scale in proportion to the reciprocal of 
the produced charged particle multiplicity, the scaling is not perfect, and
the quantity $\la N\ra\nu_{+-,dyn}$ exhibits a small dependence 
on collision centrality, which suggests the two-particle correlations 
may be modified in central collisions relative to peripheral
collisions.

A comparison of our measurement with thermal model 
predictions ~\cite{Jeon99,Jeon00,Koch02} appear to indicate fluctuations at a level
that might be expected if the Au+Au system behaved like a resonance gas. 
Although the size of the fluctuations is significantly larger than expected
in that work for a quark-gluon gas, limitations of the model used prevent
a conclusion on the existence or non-existence of a quark-gluon plasma
phase based on these results.
 
Finally, we report the first measurement of net-charge dynamical fluctuations
of identified pions, kaons, and protons. Pions exhibit dynamical fluctuations 
slightly larger than the values obtained with our inclusive measurement. 
Kaons and protons are found to exhibit dynamical fluctuations that are
2 to 4 times larger than those observed for all charged particles.
However, the lower production multiplicities of these particles may
imply the dynamical fluctuations are dominated by charge conservation effects. 
Further data are needed to assess whether the dynamical fluctuations of kaons (protons) 
significantly exceed the minimal values constrained by strangeness (baryon) 
charge conservation.

{\bf Acknowledgments:} 

We wish to thank the RHIC Operations Group and the RHIC Computing Facility
at Brookhaven National Laboratory, and the National Energy Research 
Scientific Computing Center at Lawrence Berkeley National Laboratory
for their support. This work was supported by the Division of Nuclear 
Physics and the Division of High Energy Physics of the Office of Science of 
the U.S. Department of Energy, the United States National Science Foundation,
the Bundesministerium fuer Bildung und Forschung of Germany,
the Institut National de la Physique Nucleaire et de la Physique 
des Particules of France, the United Kingdom Engineering and Physical 
Sciences Research Council, Fundacao de Amparo a Pesquisa do Estado de Sao 
Paulo, Brazil, the Russian Ministry of Science and Technology, the
Ministry of Education of China, the National Natural Science Foundation 
of China, Stichting voor Fundamenteel Onderzoek der Materie, the 
Grant Agency of the Czech Republic, Department of Atomic Energyof India, 
Department of Science and Technology of India, Council of Scientific 
and Industrial Research of the Government of India, and the Swiss National 
Science Foundation.

\end{document}